\documentclass[aps,prl,twocolumn,amsmath,amssymb,superscriptaddress,floatfix]{revtex4-1}
\usepackage{graphicx}
\usepackage{dcolumn}
\usepackage{bm}
\usepackage{amsfonts}
\usepackage{amsmath}
\usepackage{ulem}
\usepackage{color}

\begin{document}

\title{Blockade and Counterflow Supercurrent in exciton-condensate Josephson junctions} 

\author{Fabrizio Dolcini}
\affiliation{NEST-CNR-INFM and Scuola Normale Superiore, I-56126 Pisa, Italy}
\affiliation{Dipartimento di Fisica del Politecnico di Torino, I-10129 Torino, Italy}
\author{Diego Rainis}
\affiliation{NEST-CNR-INFM and Scuola Normale Superiore, I-56126 Pisa, Italy}
\author{Fabio Taddei}
\affiliation{NEST-CNR-INFM and Scuola Normale Superiore, I-56126 Pisa, Italy}
\author{Marco Polini}
\affiliation{NEST-CNR-INFM and Scuola Normale Superiore, I-56126 Pisa, Italy}
\author{Rosario Fazio}
\affiliation{NEST-CNR-INFM and Scuola Normale Superiore, I-56126 Pisa, Italy}
\author{A.H. MacDonald}
\affiliation{Department of Physics, University of Texas at Austin, Austin, Texas 78712, USA}

\begin{abstract}
We demonstrate that perfect conversion between charged supercurrents in superconductors and neutral 
supercurrents in electron-hole pair condensates is possible via a new   Andreev-like scattering mechanism.
As a result, when two superconducting circuits are coupled through a bilayer exciton condensate, the superflow in both layers is drastically modified. Depending on the phase biases the supercurrents can be completely blocked or exhibit  perfect drag. 
\end{abstract}

\maketitle

{\it Introduction} ---  The terms {\it superconductivity} and {\it superfluidity} refer to 
dissipationless flow in charged and neutral systems, respectively.  
Superfluid exciton condensates, in which macroscopic phase coherence is established among 
pairs composed of 
electrons and holes in different bands~\cite{Blatt_pr_1962,keldysh_jept_1968}, 
have been realized only recently. 
Signatures of exciton condensation  have been reported in quantum Hall bilayers~\cite{spielman_prl_2000}, in which  
electrons and holes are located in two separate two-dimensional electron layers~\cite{lozovik_jept_1975}, and 
in optically-excited exciton~\cite{butov_jphys_2007}
and exciton-polariton~\cite{exciton_polariton} cold gases. 
When the two layers of a 
bilayer exciton condensate (EC) are contacted separately, it 
can exhibit remarkable transport anomalies~\cite{spielman_prl_2000,tutuc_PRL_2004,Tiemann_NJP} 
associated with its neutral supercurrents~\cite{su_naturephys_2008}. 
These properties provide an appealing platform 
for spectacular electrical effects in EC-superconductor hybrid systems
in which the charged superconducting order parameter interfaces with the neutral EC order parameter. 
In this Letter we demonstrate that when two superconducting circuits are coupled through a bilayer EC,
the superflow in both layers is drastically altered.
If the same phase bias is applied to both junctions, no Josephson current can flow through the system,
a phenomenon we refer to as {\it exciton blockade}.  When a phase bias is applied to only one layer, on the other hand, 
it induces a {\it superdrag} counterflow supercurrent of the same magnitude in the unbiased layer.

\begin{figure}[t]
\centering
\includegraphics[width=0.70\linewidth]{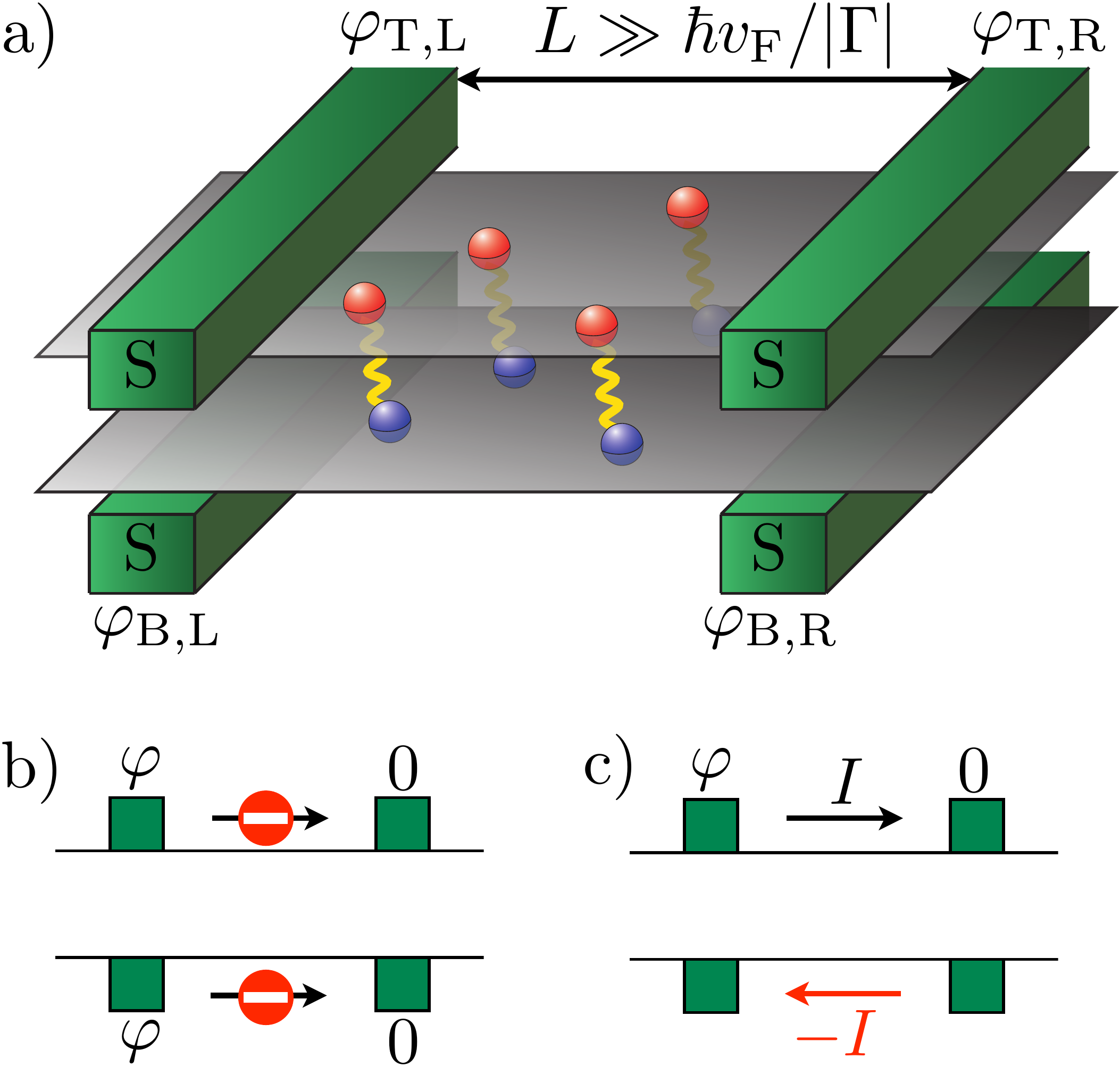}
\caption{\label{fig:one} (Color on line) Sketch of a superconductor-exciton-condensate-superconductor system.
Panel a) A double-layer exciton condensate is contacted with four superconducting leads. Panel b) When the applied phase biases in the top and bottom layers are identical ($\varphi$ in this cartoon) no current can flow when the length of the junction $L$ is much larger than the exciton-condensate coherence length $\hbar v_{\rm F}/|\Gamma|$. In this case the Josephson currents are in the ``exciton blockade" regime. Panel c) When a phase bias $\varphi$ is applied to the top layer {\it only}, a supercurrent $I$ flows. In the limit $L \gg \hbar v_{\rm F}/|\Gamma|$ a supercurrent~$-I$ is dragged in the bottom layer in a perfectly {\it frictionless} manner: in this case one has a ``perfect" drag or ``superdrag" of Josephson currents.}
\end{figure}

In order to explore the physics of conversion between EC and Cooper-pair supercurrents, 
we consider the superconductor-EC-superconductor (S-EC-S) setup sketched in Fig.~\ref{fig:one}.
Two closely-spaced layers, 
assumed to host an EC, are independently contacted to four superconducting electrodes.  The electrodes in each layer 
are separated by a distance $L$ much larger than the exciton coherence length, and an independent phase bias is applied to the top and bottom contacts. In the presence of these biases, Josephson currents flow through 
the double layer. Because the EC is gapped, only dissipationless counterflow can contribute to the Josephson current when $L$ is long.  The EC and the dissipationless nature of its counterflow supercurrent can therefore 
be revealed by a purely coherent {\it equilibrium} measurement when contacted by superconducting electrodes. 

\begin{figure}
\centering
\includegraphics[width=0.60\linewidth]{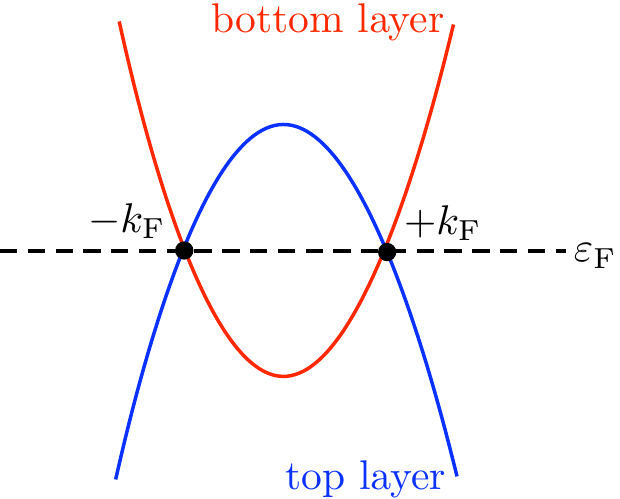}
\caption{(Color on line) Schematic electronic band structure of a gated bilayer in the absence of electron-hole coupling.
\label{fig:two}}
\end{figure}

{\it Exciton blockade and superdrag} --- All important features of the physics we want to describe are captured 
by the simple one-dimensional (1D) model that we now discuss in explicit detail. 
Later we provide arguments supporting the general validity of the conclusions we reach. 
The four superconducting leads in Fig.~\ref{fig:one} are in an equilibrium configuration, characterized by the same Fermi level 
$\varepsilon_{\rm F}$. In each layer the electron filling is controlled by a gate voltage. In our calculations we assume that the two layers are oppositely gated, so that one layer (the top layer, say) is ``hole-doped" and the other (the bottom layer) is ``electron-doped", as depicted in Fig.~\ref{fig:two}. Notice that states near the right Fermi point $+k_{{\rm F}}$  are right-movers (left-movers) for the bottom (top) layer, whereas states near the left Fermi point $-k_{{\rm F}}$ are left-movers (right-movers) for the bottom (top) layer.  Let $\Psi_{\alpha \sigma}$ be the field operator describing an electron in the $\alpha=$ top (T)/bottom (B)~$=\pm$ layer with spin $\sigma = \uparrow, \downarrow$. We have applied the Bogoliubov-de Gennes approach~\cite{tinkham_book,degennes_book} to model our S-EC-S hybrid system: the Hamiltonian $\hat{\cal H}$ of the system is
${\hat {\cal H}} =  \int^{\infty}_{-\infty}dx~\hat{\Psi}^\dagger(x) {\cal H}(x) \hat{\Psi}(x)
$, where energy is measured with respect to the equilibrium Fermi level $\varepsilon_{\rm F}$. Here  $\hat{\Psi} = (\Psi^{}_{{\rm T} \uparrow}, \Psi^{}_{{\rm B} \uparrow}, \Psi_{{\rm T} \downarrow}^\dagger, \Psi_{{\rm B} \downarrow}^\dagger  )$ 
and ${\cal H}(x)$ is a 4$\times$4 matrix  
\begin{equation}\label{ham}
   {\cal H}(x) =  
   \left(
    \begin{array}{cccc}
     {\displaystyle -\frac{ \hbar^2 \partial^2_x}{2m}} &  \Gamma(x)  & \Delta_{\rm T}(x) & 0 \\
      \Gamma^{*}(x) & {\displaystyle  \frac{ \hbar^2 \partial^2_x}{2m}}  & 0 & \Delta_{\rm B}(x)\\
     \Delta^{*}_{\rm T}(x) & 0   & {\displaystyle \frac{ \hbar^2 \partial^2_x}{2m}} & -\Gamma^{*} (x)\\
      0 & \Delta^{*}_{\rm B}(x)  & -\Gamma(x) & {\displaystyle -\frac{ \hbar^2 \partial^2_x}{2m}} 
    \end{array}
    \right)~.
\end{equation}
It contains single-particle band-kinetic-energy terms for each layer, intra-layer terms containing the superconducting order parameter~\cite{tinkham_book} 
$
	\Delta_\alpha  \propto \langle \Psi_{\alpha \downarrow} \Psi_{\alpha \uparrow} \rangle 
$
and inter-layer terms containing the EC order parameter~\cite{su_naturephys_2008} 
$
	\Gamma  \propto \langle \Psi^\dagger_{\rm B \sigma} \Psi^{}_{\rm T \sigma}  \rangle 
$.
Both order parameters vary spatially along the current-flow ($\hat{\bm x}$) direction. 
Since details of order parameter behavior near the interfaces are irrelevant for our purposes, we can 
assume that $\Delta_{\alpha}$ are non-vanishing only in the left ({\rm L}) and right ({\rm R}) electrodes. For simplicity we also assume the same amplitude $|\Delta |$ in all the electrodes, whereas the phases $\varphi_{\alpha, {\rm L}/{\rm R}}$ are allowed to differ. 
More explicitly, we take 
$
	\Delta_{\alpha}(x) = |\Delta |e^{i\varphi_{\alpha, {\rm L}}} \Theta(-x) +|\Delta |e^{i\varphi_{\alpha, {\rm R}}} \Theta(x-L)
$, 
where the origin $x=0$ is chosen at the left layer-electrode interfaces and $\Theta$ is the Heaviside step function.  
 In contrast, the EC order parameter is taken non-vanishing in the double-layer region, {\it i.e.} for $0 < x < L$.  
 Although its amplitude $|\Gamma|$ can 
be taken as constant, it is essential to allow for phase variation in order to account for condensate counterflow currents.
In 1D current conservation   implies linear phase variation so that 
$\Gamma$ has the form
\begin{equation}\label{ecorderp}
	\Gamma (x) = |\Gamma| e^{i\gamma_0 + 2iqx}~.
\end{equation}

When phase biases are applied to the four electrodes, supercurrents flow in both layers. 
The EC weak-link supports two contributions to the Josephson current.
The {\it quasiparticle} channel contribution, in which  Cooper pairs propagate 
by the virtual excitation of quasiparticles in the double layer, is present in ordinary weak links. 
In the present case, however, it is exponentially 
suppressed when $L \gg \hbar v_{\rm F}/|\Gamma|$ ($v_{\rm F}$ being the Fermi velocity) because of the gap in the
quasiparticle excitation spectrum of the EC. Much more interesting is the new contribution to the current which derives from the 
conversion of supercurrent into superfluid excitonic current.  It can be visualized as a {\it correlated Andreev 
reflection}~\cite{footnote_andreev} in which an electron and hole (in different layers) enter the EC and propagate without dissipation to the other end of the double layer.  There a similar process occurs to convert the exciton current back into a Cooper-pair current. This 
process  survives also in the {\it long-junction} limit and it leads to a number of spectacular 
effects, as we shall show below. 

In the long junction limit, $|\Delta|, |\Gamma| \gg \hbar v_{\rm F}/L$, the critical current does not depend on the magnitude of either 
order parameter. The mathematical description of the long-junction limit is simplified if we also assume that $|\Delta| \gg |\Gamma|$, with no physically relevant consequences for the main results. Indeed, for energies much smaller than $|\Delta|$ the layer spectra can be linearized~\cite{maslov_prb_1996}, and the electron field operators can be written as
$\Psi_{\alpha \sigma}(x)=e^{i k_{{\rm F}} x} \Psi_{\alpha \sigma +}(x)+ e^{-i k_{{\rm F}} x} \Psi_{\alpha \sigma -}(x)$, 
where $\Psi_{\alpha \sigma \pm}(x)$ are slowly varying fields related to the Fermi points $\pm k_{{\rm F}}$. Furthermore,  
the presence of the superconductors can be accounted for by boundary conditions at the contacts, such as 
$\Psi_{\alpha \downarrow (\uparrow ) +}(0) =  \pm \, \alpha\, i~e^{i\varphi_{\alpha, L}} \Psi^\dagger_{\alpha \downarrow (\uparrow) -}(0)$ at the left interfaces. Similarly for the right interfaces located at $x=L$. In this way the problem is reduced to the evaluation of the average of the current operator over the equilibrium state of a system in which the fields satisfy these boundary conditions {\it and} the exciton order parameter (\ref{ecorderp}) exhibits a space-dependent phase winding.  It follows that the EC winding wavevector $q$ must satisfy 
\begin{equation}
\label{eq:q-def}
 q=	\frac{\varphi_{\rm T}  - \varphi_{\rm B} - 2 \pi J}{4L} 
\end{equation}
with $J$  an integer. Here $\varphi_{\alpha} \equiv \varphi_{\alpha, {\rm L}}- \varphi_{\alpha, {\rm R}}$ is the phase bias in layer $\alpha$.
Minimization of the total energy fixes  $J$ to be the closest integer to $(\varphi_{\rm T}  - \varphi_{\rm B})/(2\pi)$, and the offset phase  to
be $\gamma_0 =  (\varphi_{{\rm T}, {\rm L}}-\varphi_{{\rm B}, {\rm L}})/2$.
We find that, at zero temperature, the supercurrents in top and bottom layers flow in opposite directions. 
Explicitly, they exhibit a sawtooth form,
\begin{equation}\label{eq:jodrag}
I^{(0)}_{{\rm T}/{\rm B}}  =  \pm~\frac{e v_{\rm F} }{ 2 \pi L }(\varphi_{\rm T}-\varphi_{\rm B})~,
\end{equation}
where $\varphi_{\rm T}-\varphi_{\rm B}$ is defined modulo $2\pi$. The magnitude of the currents depends on {\it the difference} $\varphi_{\rm T}-\varphi_{\rm B}$ {\it between the phase biases} $\varphi_{\rm T}$ and 
$\varphi_{\rm B}$ in the two layers. 

Eq.~(\ref{eq:jodrag}) is the  main result of this Letter and has several interesting physical implications:
i) $\varphi_{\rm T} = \varphi_{\rm B}$ (parallel flow). 
When the same phase biases are applied to the two junctions no supercurrents can flow through the EC. In this case the Josephson currents experience an ``exciton blockade";
ii) $\varphi_{\rm T} = -\varphi_{\rm B}$ (counterflow). In this case the Josephson current flowing through the EC is maximal, with a 
critical value equal to the critical current of a ballistic one-channel  superconductor - normal metal - superconductor (S-N-S) junction; 
iii)  $\varphi_{\rm T} = \varphi$ and $\varphi_{\rm B}= 0$ (superdrag). When current flows in one layer due to a phase bias in that layer, a current equal in magnitude but opposite in direction flows in the other layer. This is a consequence of perfect conversion of exciton current into supercurrent.  Eq.~(\ref{eq:jodrag}) can then be seen as a perfect drag effect for the supercurrent.

The existence of a dissipationless (counterflow) channel also has a spectacular impact on the temperature dependence of the  critical current. Indeed we notice that the ground-state current (\ref{eq:jodrag}) has the same length dependence as that in a S-N-S junction~\cite{IcSNS-T=0}.
At finite temperature it is possible to show that, in the regime 
$\hbar v_{\rm F}/L \ll k_{\rm B} T \ll |\Gamma|$, the critical current in the S-EC-S is
\begin{equation}\label{cur-finite-T}
I_{{\rm T}/{\rm B}} = \pm \frac{2 e v_{\rm F}}{\pi} q \left[ 1-  \sqrt{2 \pi  \beta|\Gamma|}~\frac{\sinh(qL_{\rm th})}{q L_{\rm th}}~e^{-\beta |\Gamma|}\right]~,
\end{equation}
where $\beta = 1/(k_{\rm B} T)$ and $L_{\rm th} =\hbar v_{\rm F}/(k_{\rm B} T)$ is the thermal length. Note that 
the second term in square brackets in Eq.~(\ref{cur-finite-T}) is $\propto \exp(-\beta|\Gamma|)$: thus, as long as thermal fluctuations are dominated by the excitonic gap, the ground-state current is essentially unaffected by thermal fluctuations. 
This is another important result of this Letter. Notice that this occurs even when the thermal length 
$
L_{\rm th}
$ 
is smaller than the length $L$ of the junction. This is in striking contrast with the case of a 
S-N-S junction (or with the case of two decoupled layers), {where the critical current $I_{\rm c}$ is exponentially suppressed~\cite{IcSNS}, {\it i.e.} $I_{\rm c} \propto \exp{(-L/L_{\rm th})}$ for $L_{\rm th} \ll L$}, due to thermal decoherence affecting a single Andreev-reflection process. In the presence of the EC, Andreev processes coherently occurring in the two layers transform Cooper pairs into electron-hole pairs of the EC, which are protected from thermal decoherence by the excitonic gap. Thus in the temperature window 
$
\hbar v_{\rm F}/L \ll k_{\rm B} T \ll |\Gamma|
$ 
the EC counterflow channel is responsible for an {\it exponential enhancement} of the critical current {with respect to the ordinary S-N-S case mentioned just above}.

{\it Discussion} --- The only crucial assumption we made in the previous derivation is that the length $L$ of the junction is much larger than the EC coherence length $\hbar v_{\rm F}/|\Gamma|$. For these reasons, the physical results obtained are not restricted to the specific  1D model discussed above. The dependence of the Josephson current  on the difference $\varphi_{\rm T}-\varphi_{\rm B}$ can be deduced from quite general arguments. Current conservation indeed implies that the supercurrent can be evaluated in the bulk of the layers, where the supercurrent is purely carried by the EC, provided that the junction is long enough ($L\gg \hbar v_{\rm F}/|\Gamma|$). We also emphasize that, due to the charge neutrality of the EC order  parameter~$\Gamma$,  the supercurrents in the two layers are equal in magnitude and opposite in sign. In particular, 
the current is proportional to~$q$, {\it i.e.} the phase winding in Eq.~(\ref{ecorderp}). 
The evaluation of the current thus reduces to the  determination of the dependence of~$q$ on the superconducting phase biases $\varphi_{\rm T}$ and $\varphi_{\rm B}$. Let us assume that the system is translationally-invariant in the transverse direction ${\hat {\bm y}}$, so that the order parameters $\Delta_\alpha({\bm r})$ and $\Gamma({\bm r})$ depend on the longitudinal ${\hat {\bm x}}$-direction only. By applying the transformation 
$
\Psi_{\alpha \sigma}({\bm r}) = \exp{\{i [\varphi_{\alpha, {\rm L}} + (\varphi_{\alpha, {\rm R}} - \varphi_{\alpha, {\rm L}})x/L]/2\}} 
{\widetilde \Psi}_{\alpha \sigma}({\bm r})
$, 
the superconducting phase biases $\varphi_{\alpha, {\rm L}/{\rm R}}$ can be gauged away. The price to pay is twofold. 
Firstly, an effective vector potential ${\bm A}_\alpha = (\Phi_0 \varphi_\alpha/ L)~{\hat {\bm x}}$ appears in each intra-layer Hamiltonian (here $\Phi_0 = h / 2e$ is the quantum of flux, associated with elementary charge $2e$ of the superconducting order parameter $\Delta_\alpha$). 
Secondly, the EC order parameter describing the inter-layer coupling transforms into 
\begin{equation}\label{eq:final_gamma}
{\widetilde \Gamma}(x) = |\Gamma| \exp{\left[2i \left(q - \frac{\varphi_{\rm T}-\varphi_{\rm B}}{4L}\right) x\right]}~.
\end{equation}
By observing that the energy scales characterizing the intra-layer and the inter-layer terms are $\hbar v_{\rm F}/L$ and $|\Gamma|$, respectively, it is straightforward to realize that for a long junction ($L\gg \hbar v_{\rm F}/|\Gamma|$) the inter-layer terms play the major role in determining the equilibrium configuration. Energy minimization implies that the argument of the exponent proportional to $x$ in (\ref{eq:final_gamma}) vanishes, since in the new gauge the system is effectively phase unbiased. This fixes the winding $q$ to be the one defined in Eq.~(\ref{eq:q-def}), and proves that the currents just depend on $\varphi_{\rm T}-\varphi_{\rm B}$. The current-phase relationship is always of the sawtooth form.
We stress that the above
argument {\it does not depend} on the details of the experimental setup. In particular, it applies independently of the specific (parabolic or linear) energy-momentum dispersion relation of the intra-layer kinetic Hamiltonian. Furthermore, it also holds if the contacts with the superconducting electrodes are not ideal and when their transparencies are different in the top and bottom layers. In the specific case of 2D layers with width $W$ and highly-transparent contacts, the current is still given by Eq.~(\ref{eq:jodrag}) provided that it is multiplied by the number $k_{\rm F} W/4$ of transverse channels~\cite{IcSNS-T=0,IcSNS,S2degS}.

The unique properties of the conversion of EC currents into charged supercurrents 
can be exploited for a number of possible 
applications.     
As an example we discuss a configuration  realized 
by closing the two superconducting electrodes
contacted to the (say) top layer into a ring-shaped rf-SQUID geometry, so that the 
phase difference $\varphi_{\rm T}$ is directly related to the magnetic flux $\Phi_{\rm T}$ 
by $\varphi_{\rm T}=2 \pi \Phi_{\rm T}/\Phi_0 +2 \pi n$.
In response to a magnetic field, an induced  Josephson current $I_{\rm T}$ flows in the top layer  and, according to 
Eq.~(\ref{eq:jodrag}), an opposite current $I_{\rm B} = - I_{\rm T}$ flows in the bottom layer. 
Whenever the magnetic field changes the flux by a 
fluxon, the currents in both layers are reversed. Current sign switches detected in the bottom layer count the 
fluxons present in the top layer ring. If the magnetic flux is generated by a monotonic analog input signal, the system effectively converts it into a sum of current switch pulses, {\it i.e.} to a digital signal.  The system is therefore an analog-to-digital converter. A generalization to non-monotonic input signals can easily be achieved by using two double junctions.

One important obstacle which presently stands in the way of observing these effects 
is the fact that equilibrium exciton condensation has so far been observed only in 
Quantum Hall (QH) bilayers at total filling factor $\nu_{\rm T}=1$.  QH systems necessarily have 
current-carrying gapless channels localized at their edges.  In a QH bar geometry 
the edge channels will alter the physics we discuss.  
Spontaneous coherence between conduction and valence band electrons in 
different semiconductor quantum well layers is however also expected~\cite{senatore} to occur
at zero magnetic field when inter-layer interactions are strong. 
There are hints that the conditions necessary for coherence 
have been realized in some recent~\cite{semiconductor_coherence} semiconductor bilayer experiments.
Graphene bilayer systems~\cite{graphene_review_1} are just starting to be examined for 
coherence effects and have a number of potentially important advantages,
as pointed out recently by several researchers~\cite{min_prb_2008,joglekar_prb_2008,lozovik_jept_2008}. 
Because they are gapless and atomically 2D, the field-driven carrier densities that 
can be achieved are much larger than in the semiconductor case.  Weaker dielectric screening and 
linearly dispersive conduction and valence bands help to increase both interaction and 
disorder energy scales.  Finally, graphene bands are nearly perfectly particle-hole 
symmetric, guaranteeing the nearly perfect nesting between conduction and valence band Fermi surfaces which favors the coherent state.
Progress~\cite{schmidt_apl_2008,tutuc_arXiv_2009,eva_natphys_2009,jens_MM_2009} in the realization of 
electrically isolated double-layer graphene sheets,
either two-layers separated by a dielectric or rotated layers,  
is on-going. In the single-layer graphene case, it has already been demonstrated~\cite{graphene_weak_link_1} 
that it is possible to fabricate transparent interfaces between graphene and superconducting electrodes. 
In view of this progress our predictions are likely to be within experimental reach soon. 

{\it Acknowledgments} --- We acknowledge stimulating conversations with P. Jarillo-Herrero and financial support by CNR-INFM ``Seed Projects'' (F.D., F.T., and M.P.), Italian MIUR ``Rientro dei Cervelli" program (F.D.), Welch Foundation, NSF and SWAN NRI program (A.H.M.).


\begin{thebibliography}{77}
\bibitem{footnote_andreev} 
	Conventional ``Andreev reflection" is a coherent electron-hole conversion process 
	which occurs at the interface between a normal metal and a superconductor 
	(see Ref.~\onlinecite{tinkham_book}).
\end{thebibliography}
\end{document}